\documentclass[10pt,conference]{IEEEtran}

\usepackage{tikz-cd}
\usepackage{tabularx}
\usepackage{listings, color}
\usepackage{multirow}
\usepackage[utf8]{inputenc}
\usepackage[english]{babel}
\usepackage[T1]{fontenc}

\usepackage{caption}
\usepackage{subcaption}
\usepackage{proof}
\usepackage{algorithm2e}
\usepackage{color}
\usepackage{xcolor}

\newcommand{\term}[1]	{\emph{#1}}

\newcommand{\BD}{\begin{definition}}
\newcommand{\ED}{\end{definition}}

\newcommand{\code}[1]{\texttt{#1}}

\newif\ifnotreviewed
\newcommand{\forreview}[1]{\ifnotreviewed \textcolor{red}{#1} \else #1\fi}

\newcommand{\SPL}       {\mathcal{L}}
\renewcommand{\L}		{\SPL}

\newcommand{\config}{\rho}

\newcommand{\indexProd}[2]{#1|_{#2}}
\newcommand{\onefifty}{150\% representation}

\newcommand{\lift}[1]{#1^\uparrow}

\newcommand{\souffle}{Souffl\'{e}}
\newcommand{\lsouffle}{Souffl\'{e}$^\uparrow$}

\definecolor{codegreen}{rgb}{0,0.6,0}
\definecolor{codegray}{rgb}{0.5,0.5,0.5}
\definecolor{codepurple}{rgb}{0.58,0,0.82}
\definecolor{backcolour}{rgb}{0.95,0.95,0.92}

\lstdefinestyle{GrokStyle}{
    backgroundcolor=\color{backcolour},   
    commentstyle=\color{codegreen},
    keywordstyle=\color{magenta},
    numberstyle=\tiny\color{codegray},
    stringstyle=\color{codepurple},
    basicstyle=\footnotesize,
    breakatwhitespace=false,         
    breaklines=true,                 
    captionpos=b,                    
    keepspaces=true,                 
    numbers=left,                    
    numbersep=5pt,                  
    showspaces=false,                
    showstringspaces=false,
    showtabs=false,                  
    tabsize=2
}

\newcommand{\GM}{General Motors}
\newcommand{\EEightZ}{SPL-A}    
\newcommand{\EEightE}{SPL-B}    
\newcommand{\ENine}{SPL-C}      
\newcommand{\ENineN}{SPL-D}      
\newcommand{\EZero}{SPL-E}      
\newcommand{\EZeroT}{SPL-F}      

\begin{document}

\title{Applying Declarative Analysis to Software Product Line Models: An Industrial Study}

\author{
    \IEEEauthorblockN{Ramy Shahin}
	\IEEEauthorblockA{University of Toronto\\
		\small{rshahin@cs.toronto.edu}}
	\and
	\IEEEauthorblockN{Robert Hackman, Rafael Toledo}
	\IEEEauthorblockA{University of Waterloo\\ \small{\{r2hackma,rftoledo\}@uwaterloo.ca}}
	
	\and
	\IEEEauthorblockN{Ramesh S.}
	\IEEEauthorblockA{General Motors\\ \small{ramesh.s@gm.com}}
	\and
	\IEEEauthorblockN{Joanne M. Atlee}
	\IEEEauthorblockA{University of Waterloo\\ \small{jmatlee@uwaterloo.ca}}
	\and
		\IEEEauthorblockN{Marsha Chechik}
	\IEEEauthorblockA{University of Toronto\\ \small{chechik@cs.toronto.edu}}

}
	
\maketitle

\begin{abstract}
Software Product Lines (SPLs) are families of related software products developed from a common set of artifacts. Most existing analysis tools can be applied to a single product at a time, but not to an entire SPL. Some tools have been redesigned/re-implemented to support the kind of variability exhibited in SPLs, but this usually takes a lot of effort, and is error-prone. Declarative analyses written in languages like Datalog have been collectively \emph{lifted} to SPLs in prior work~\cite{Shahin:2019}, which makes the process of applying an existing declarative analysis to a product line more straightforward.

In this paper, we take an existing declarative analysis (\emph{behaviour alteration}) and apply it to a set of automotive software  product lines from \GM. We discuss the design of the analysis pipeline used in this process, present its scalability results, and provide a means to visualize the analysis results for a subset of products filtered by feature expression. We also reflect on some of the lessons learned throughout this project.
\end{abstract}

\section{Introduction}\label{sec:intro}
Software Product Lines (SPLs) are families of related products, usually developed together from a common set of artifacts. 
Each product configuration is a combination of features, each of which can be either present or absent in a product. Due to this combinatorial nature of features, the number of potential products grows exponentially with the size of the feature set.
The high level of configurability within an SPL is usually desired. However, analysis tools (such as syntax analyzers, type checkers, model checkers, static analysis tools) are typically designed to work on a single product, not a whole SPL. Applying an analysis to each product separately is usually infeasible for non-trivial SPLs because of the exponential number of products~\cite{Liebig:2013}.

One class of analyses is those written in declarative languages (e.g., Datalog~\cite{Ceri:1989}, Grok~\cite{grok97}). An example is the \term{behaviour alteration analysis}~\cite{muscedere2019detecting}, which detects interactions between components in a software system. The first step in applying a declarative analysis to a software system is to extract the relevant facts from the system; then the logical rules of the analysis are applied to those facts, generating results. Like most program analyses, our behaviour alteration analysis can be applied only to a single software product, not an SPL.

Since all products of an SPL share a common set of artifacts, analyzing each product individually (usually referred to as \term{product-based analysis}~\cite{Thum:2014}) would involve a lot of redundant analyses. How to leverage the high degree of  commonality across products and analyze the whole product line at once, bringing the total analysis time down, is a fundamental research problem at the intersection of product-line engineering and software analysis. Different attempts have been made to \term{lift} individual analyses to run on product lines~\cite{Bodden:2013, Classen:2010, Gazzillo:2012, Kastner:2011, Kastner:2012, Midtgaard:2015, Salay:2014}. The resulting \term{variability-aware analyses}~\cite{Thum:2014}, which analyze the SPL as a whole, show significant time savings compared to the product-based analyses on the SPL's set of products. The downside is the amount of effort required to correctly lift each of those analyses. 

In previous work, we lifted a whole class of analyses as opposed to a single analysis. Specifically, we designed and implemented a \term{variability-aware} Datalog engine~\cite{Shahin:2020a} that can be used to efficiently apply an existing single-product Datalog analysis to facts extracted from a whole product line~\cite{Shahin:2019}, and we used the engine to apply a lifted pointer and taint analyses to Java product lines~\cite{Shahin:2019} and other lifted analyses to C-language product lines~\cite{Shahin:2021}. 
\forreview{Our approach applies to \textit{annotative} SPLs, in which each element in an SPL artifact is annotated with the features it belongs to.}
In this project, we leverage the variability-aware Datalog engine to lift the behaviour alteration analysis  
and we apply the analysis to six vehicle controller 
provided by \GM.
This project also considers how to visualize the analysis results, so that the engineer can easily explore the results from the perspective of different sets of products.

This paper makes the following contributions: (1) We outline the design of a pipeline for variability-aware behavior alteration analysis. (2) We present the results of applying our pipeline to a set of automotive software product lines from \GM. (3) We discuss the lessons learned throughout the project.

The rest of the paper is organized as follows. Section~\ref{sec:background} provides a background on SPLs, fact extraction, behaviour alteration analysis, and lifted declarative analyses. In Section~\ref{sec:implementation}, we present our approach to lifting behaviour alteration analysis, including configurable visualization of analysis results. In Section~\ref{sec:caseStudy} and \ref{sec:evaluation}, we present our industrial examples and the results of applying our lifted analysis to them. We discuss lessons learned in Section~\ref{sec:lessons}, present related work in Section~\ref{sec:related}, and conclude in Section~\ref{sec:conclusion}.

\section{Background} 
\label{sec:background}

\newcommand{\FA}{\code{FA}}
\newcommand{\FB}{\code{FB}}
\newcommand{\FC}{\code{FC}}
In this section, we briefly define the concepts we build upon in the rest of the paper. In particular, this includes background on software product lines, declarative analyses of relational models, the behavior alteration analysis, and visualization of variabiliy results.

\subsection{Software Product Lines}
\begin{figure*}[t]

\begin{subfigure}[]{0.49\textwidth}
	\lstset{style=GrokStyle}
    \begin{lstlisting}[language=C++,numbers=left]
    extern int GlobVar; // shared from C2
    extern bool FA; // Feature variable
    extern bool FB; // Feature variable
    ...
    class A {
        int x = 0;
        
        int updateX() {
            if (FA) {
                x = GlobVar * 2;
            
                if (FB) {
                    x++;
                } else { // !FB
                    x = (++GlobVar) * 2;
                } // FB
        }
        ...
        } // FA
    }
    ...
    \end{lstlisting}
	\caption{Component \code{C1}.}
	\label{fig:C1}	
\end{subfigure}
\hfill
\begin{subfigure}[]{0.49\textwidth}
	\lstset{style=GrokStyle}
	\begin{lstlisting}[language=c,numbers=left]
	int GlobVar = 0; // shared global
	extern bool FB; // Feature variable
	
	int foo() {
	    ...
	    if (FB) {
	        return GlobVar;
	    }
	}
	int bar() {
	    ...
	    if (GlobVar > 20) {
	        return foo();
	    }
	}
	\end{lstlisting}
	\caption{Component \code{C2}.}
	\label{fig:C2}	
\end{subfigure}
\caption{An example of a Software Product Line with features \code{FA} and \code{FB}, and  components \code{C1} and \code{C2}.}
\label{fig:splExample}
\end{figure*}

A \term{Software Product Line (SPL)} is a family of related software products, developed together from a common set of artifacts~\cite{Clements:2001}. The unit of variability in an SPL is a \term{feature}, where each feature can be either present or absent in each of the individual products. Because of the combinatorial nature of SPL features, the number of products grows exponentially with the number of features. \forreview{However, there are typically constraints among features that preclude all possible feature combinations from generating valid products.} A Feature Model~\cite{Kang:1990} captures the set of valid feature combinations.

For example, the SPL in Figure~\ref{fig:splExample} has two features, \FA~and \FB. \FA~and \FB \ are assumed to be compile-time Boolean constants, each indicating whether its corresponding feature is present or absent in the  product. Feature-specific code is guarded within conditional statements, with \term{feature expressions} (propositional formulas over features) as conditions. This example is usually classified as an \term{annotative} SPL~\cite{Thum:2014} because different program artifacts (lines of source code in this case) can be individually annotated with feature expressions. The whole program (the union of all features) is usually referred to as the \term{\onefifty} of the product line (because it is generally true that there is no valid product that includes all features, due to feature constraints).

This annotation mechanism allows assigning a given code block a \term{Presence Condition (PC)}, which specifies a feature expression denoting the set of products in which the line exists. For example, in Figure~\ref{fig:C1}, the PC of line 10 is \FA, meaning that line 10 
exists in all, and only in, products that include  feature \FA. Line 13, on the other hand, has $(\FA \land \FB)$ as a PC because both features \FA~and \FB~need to be included in a product for this line to exist. Similarly, the PC of line 15 is $(\FA \land \lnot \FB)$, indicating that this line only exists in products including feature \FA, and excluding feature \FB.

A single software product can be generated from an SPL given a \term{feature configuration}, which specifies the set of features to be included in the product to be generated. For example, in Figure~\ref{fig:C1}, the feature configuration \{\FA\} would generate a program with all code blocks except for line 13 because \FB~is not included in the configuration. The configuration \{\FA,~\FB\}, on the other hand, would generate all blocks of code except for line 15, which would only exist in products where \FA~is present and \FB~is absent. We refer to the product generated from product line $\L$ and feature configuration $\config$ as $\indexProd{\L}{\config}$.

The primary motivation behind developing a family of products together as an SPL instead of developing each product independently is to maximize reuse of common software artifacts across products, leveraging the potentially high degree of commonality among them. Different techniques of developing SPLs have been proposed and used in practice~\cite{Gacek:2001, Apel:2009, Schaefer:2010}.

A typical software development life cycle also includes the use of various tools to perform a variety of analyses of software artifacts. These include tools for bug-finding, metric generation, and performance assessment. In most cases, the tools can be applied only to one software product at a time rather than to the entire SPL. The naive approach of generating each and every product and applying an analysis tool to it individually is usually infeasible because of the exponential growth in the number of products as the number of features increases. 

\subsection{Lifted Declarative Analyses}
Several software analyses have been re-designed and implemented to support efficiently analyzing the whole SPL at once (more on this in Section~\ref{sec:related}). Those are usually referred to as \term{variability-aware analyses}, and the process of transforming a single-product analysis to an variability-aware analysis is usually referred to as \term{variability-aware lifting}~\cite{Bodden:2013, Salay:2014, Shahin:2019, Shahin:2020b}. A \term{lifted} analysis is expected to preserve the semantics of its single-product counterpart, while tracing each of the results of the analysis to the set of products to which it applies. We use the notation $\lift{f}$ to refer to a lifted version of a product-based analysis $f$.

For example, assume that we are designing an analysis that detects interference between software components through the use of global variables. Component \code{X} interferes with component \code{Y} if \code{X} writes to some global that is read by \code{Y}. The example SPL in Figure~\ref{fig:splExample} shows two components, \code{C1} and \code{C2}, both of which access a global variable \code{GlobVar}. Component \code{C2} (Figure~\ref{fig:C2}) reads the value of \code{GlobVar} in lines 7 and 12, but the first read exists only in products that include feature \FB, while the second exists in all products. Component \code{C1} (Figure~\ref{fig:C1}) reads from \code{GlobVar} in lines 10 and 15, and it also writes to \code{GlobVar} in line 15. The presence condition of line 15 is $(\FA \land \lnot \FB)$, which means that \code{C1} interferes with the first read of \code{C2} (line 7); and if \FB~is included in the product, it might also affect the control-flow of function \code{bar} depending on the value of \code{GlobVar} at line 12.

A variability-aware analysis is expected to report the results of applying the corresponding product-based analysis to each product configuration, annotating each of the results with its correct product configuration. For the example in Figure~\ref{fig:splExample}, it is not enough to report that \code{C1} interferes with \code{C2}. Instead, the analysis needs to specify the product configuration(s) in which this interference occurs (\{\FA\} in this case).

Formally, given a product line $\L$ and a lifted analysis function $\lift{f}$, applying $\lift{f}$ to $\L$ and restricting the results to a single product denoted by product configuration $\config$ should be the same as applying the original analysis $f$ to the set of artifacts belonging only to the product configuration $\config$~\cite{Shahin:2020b}. 
%

Instead of re-implementing a given analysis to make it variability-aware, another approach is to lift the language in which the analysis has been implemented. This has the advantage of not having to modify the original product-based analysis at all, and if many analyses are written in the lifted language, they all get lifted for free. For example, analyses written in Datalog have been collectively lifted~\cite{Shahin:2019} by extending the Datalog language with optional presence condition annotations at the fact level, and implementing a variability-aware fact inference algorithm in the \lsouffle~(lifted \souffle) Datalog engine~\cite{Shahin:2020a}.

\subsection{Behaviour Alteration Analysis}

%
%

A \term{behaviour alteration}~\cite{Muscedere:2019} is a form of
data-flow \term{component interaction} 
that occurs when a change to a variable value made in one component alters the behaviour of another component. Such an analysis is useful in large component-based systems, where an engineering team knows its components well, but does not know all of the ways in which actions taken in its components can affect the behaviours of other teams' components.
The specific instance of behaviour alteration used in this paper is (1) an assignment made in component \code{C1} to a variable \code{v}, (2) whose value impacts other variables through variable assignments, and impacts other components through parameter passing; until (3) a variable \code{x} whose value has been influenced by the modified value of \code{v} is used in
the decision condition of some control structure (i.e., an \texttt{if}, \texttt{for},
\texttt{while}, or \texttt{switch} statement) in another component \code{Cn} that (4) guards a function call. Thus, the analysis looks for a data flow from a variable assignment in one component to a control structure in another component, where the control-structure's statement block includes a function call. 
Figure~\ref{fig:splExample} gives a simple example
where the write to \texttt{GlobVar} in line 15 of component \code{C1} could affect whether or not the function
\texttt{bar} calls the function \texttt{foo} in component \code{C2}.

The analysis operates on extracted ``facts" about C/C++ source code, rather than operating on the code itself, to enable the analysis to scale to large software systems. Specifically, a fact extractor Rex~\cite{Muscedere:2019}, based on the \textit{Clang++}~\cite{clang} open-source compiler, parses C/C++ source-code files, generates abstract syntax trees (ASTs), and extracts facts of interest from the AST into an in-memory hierarchical graph. Source-code entities such as variable declarations and function declarations are the nodes of the graph; and relations such as variable assignments (in which one variable is used in the  assignment expression for another variable), function calls, and containment (of variable declarations within functions, function declarations within files, components comprising files) are the edges of the graph.  Additional information about the nodes and edges are recorded as associated attributes. Rex outputs the resulting graph as a collection of facts (called a \term{factbase}) about source-code entities, their relations, and their respective attributes represented as three-tuples (triples) in the Tuple-Attribute (TA) language~\cite{holt1997introduction}. 

\begin{figure}[t]
    \centering
    \lstset{style=GrokStyle}
    \begin{lstlisting}[language=C++]
transVarWrite(v0, v1) :- varWrite(v0, v1).
transVarWrite(v0, v2) :- varWrite(v0, v1), 
                         transVarWrite(v1, v2).

behAlter(f0, f1) :- write(f0, v0), 
                    transVarWrite(v0, v1), 
                    varInfFunc(v1, f1),
                    cFunction(f0, c0), 
                    cFunction(f1, c1), 
                    c0 != c1.
\end{lstlisting}
    \caption{\forreview{Datalog
    definition for detecting symptoms of behaviour alterations.}}
    \label{fig:HS5}
\end{figure}
\forreview{
Figure~\ref{fig:HS5} shows a definition (simplified for presentation purposes) for detecting symptoms of behaviour alteration between software components, expressed as Datalog rules applied to facts extracted from a program.  Lines 1-3 compute the transitive closure of the \texttt{varWrite} relationship, thereby finding all data-flows in which one variable is used in the assignment expression of another variable (including parameter assignments).  Lines 5-10 define behaviour alteration as a data-flow (\texttt{transVarWrite}) that starts with a variable assignment  (\texttt{write}) in function 
\texttt{f0}, and ends with function \texttt{f1} whose invocation is influenced by a variable value (\texttt{varInfFunc}). As we are only interested in behavior alterations crossing component boundaries, we exclude intra-component alterations (lines 8-10).}

Running the analysis on the facts extracted from the code in Figure~\ref{fig:splExample} yields that the function \texttt{updateX} in component \code{C1} may influence whether the function \texttt{foo} in \code{C2} is called:
(i) \texttt{write} relationship from function \texttt{updateX} to variable \texttt{GlobVar} (line 15 in \code{C1}); (ii) \texttt{varWrite} relationship from the variable \texttt{GlobVar} to itself (line 15 in \code{C1}); (iii) \texttt{varInfFunc} relationship from the variable \texttt{GlobVar} to the function \texttt{foo} (lines 12-13 in \code{C2}).

\subsection{Visualizing edge groups}
In the current work, we visualize the results of lifted behaviour alternation analysis as paths in the hierarchical graph that represents the fact base. Thus, the above example of behaviour alteration would be a path comprising (1) a \texttt{write} edge from function node \texttt{updateX} to variable node \texttt{GlobVar}, (2) a \texttt{varWrite} edge that loops from variable node \texttt{GlobVar} to itself, and (3) a \texttt{varInfFunc} edge from variable node \texttt{GlobVar} to function node \texttt{foo}. 


It is common in graph visualization to identify a group of edges via some visual aspect, such as styling, position, thickness, shape, or colour~\cite{vehlow2017visualizing}. Approaches that employ colour assign a different colour to each edge group or type of edge, thereby improving the readability of the graph~\cite{gansner2006improved}. Edges that belong to more than one group can be represented by multiple edge instances, each assigned a single colour representing one of the edge's groups~\cite{yoghourdjian2015high} or can be represented by a multi-coloured edge~\cite{dai2013vistruclizer} that comprises the colours associated with all of the edge's groups.
Such techniques have been used to support the visual analysis of graphs representing social networks~\cite{dai2013vistruclizer, henry2006matrixexplorer} and source code attributes~\cite{erdemir2011quality, abuthawabeh2013immv}.



\section{The Analysis Pipeline}
\label{sec:implementation}
\begin{figure*}[t]
	\centering
	\includegraphics[width=0.9\textwidth]{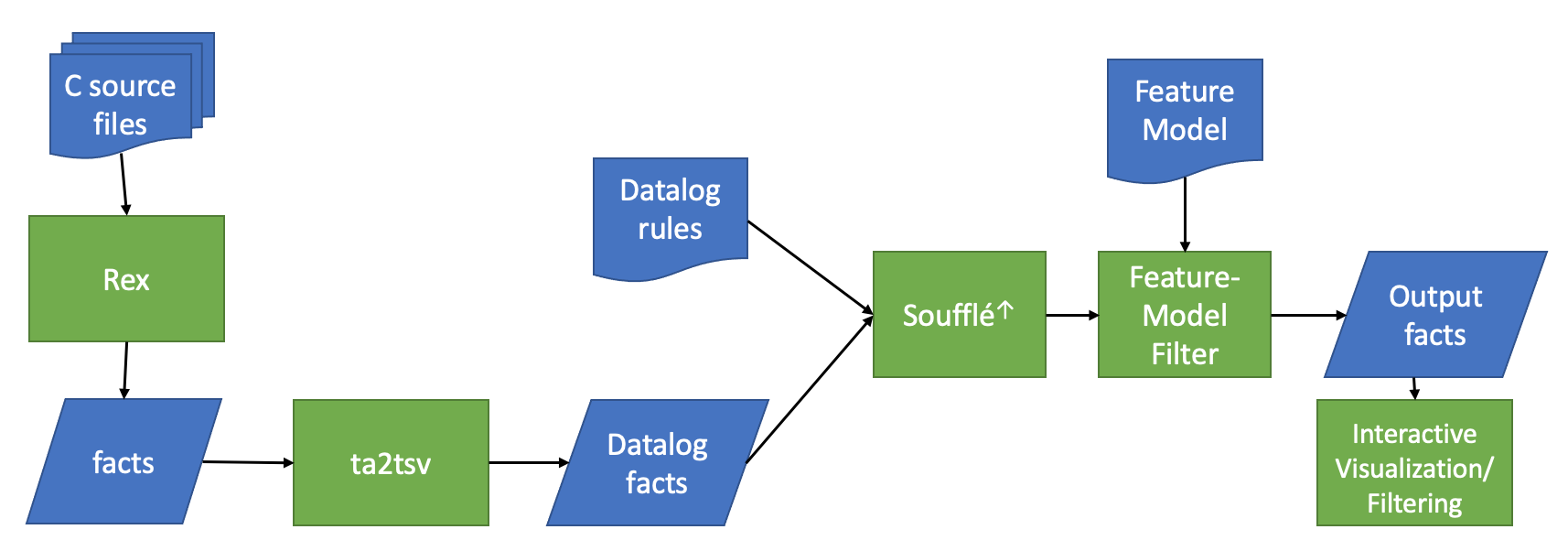}
	\caption{End-to-end fact extraction and analysis pipeline.} 
	\label{fig:pipeline}
\end{figure*}
We implemented an end-to-end pipeline for extracting a product line model from source code, analyzing it, and interactively visualizing the results.
The analysis pipeline integrates components used in previous projects~\cite{Shahin:2019, Muscedere:2019}, together with some adapter components for converting data from one format to another. The overall pipeline design is shown in Figure~\ref{fig:pipeline}.

An SPL model is extracted from C-language source files using a new variability-aware version of \emph{Rex}, which extracts syntactic facts about the source files (e.g., variable declarations, variable assignments, function declarations, function calls) and annotates a fact with a presence condition (PC) if the fact relates to code that is present in a subset of products. Facts and their presence conditions are extracted in \emph{Tuple-Attribute (TA)} format~\cite{holt1997introduction}, which are then converted to Datalog fact format using the \emph{ta2tsv} adapter component. 


The behaviour alteration analysis is expressed as a collection of Datalog rules. Datalog facts and rules are fed as inputs to \lsouffle, a variability-aware Datalog engine. The output of \lsouffle~can be optionally filtered using a Feature Model, removing those facts that do not belong to any of the valid products of the SPL. We describe components of this pipeline below.

\subsection{Variability-aware Fact Extraction}
\begin{figure*}[t]
	\centering
	\includegraphics[width=\textwidth]{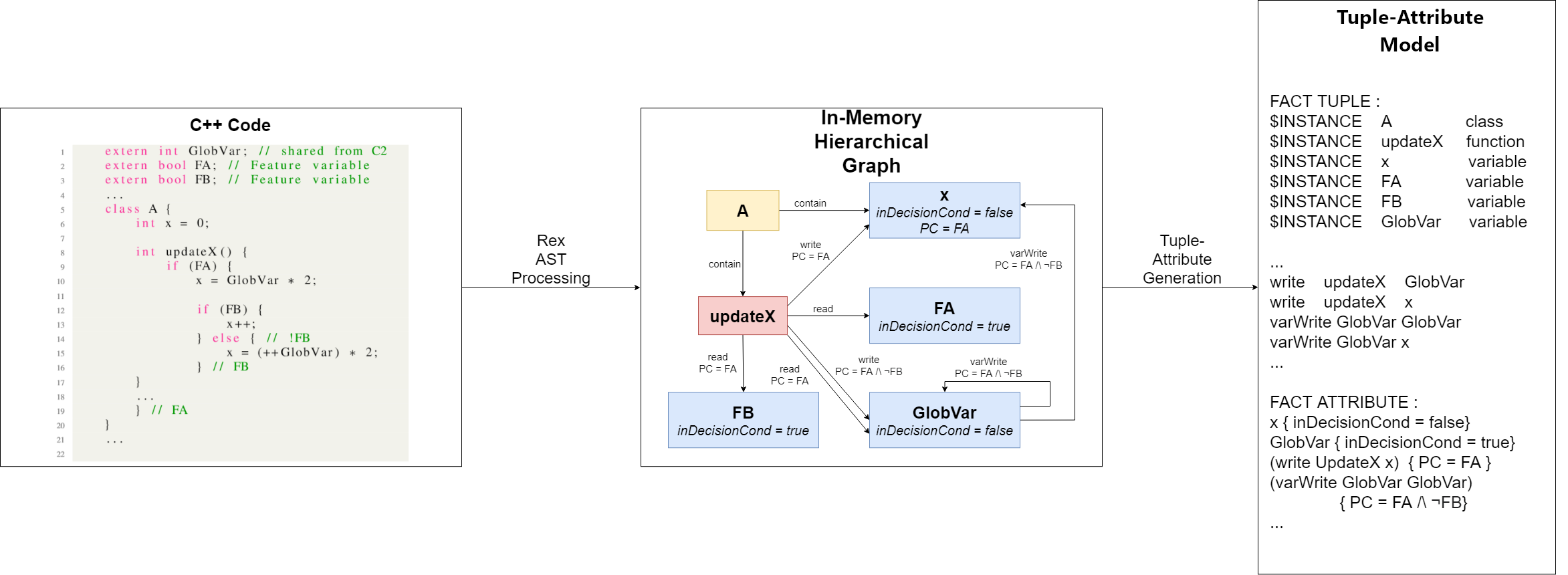}
	\caption{The fact extraction of component \code{C1}.}
	\label{fig:factExtraction}
\end{figure*}


In order to support analysis of SPL models, we developed a variability-aware version of
Rex that annotates entities and relationships with their presence conditions. A Rex user can specify, by type and naming convention, 
which program variables are to be considered feature variables to be used in presence conditions (e.g., only constant global \texttt{bool} or
\texttt{enum} type variables). Variability-aware Rex keeps track of all relevant predicates currently in effect
while walking the AST and uses that information to annotate the model as it is extracted.

Figure~\ref{fig:factExtraction} gives an overview of the Rex extraction process of the component \code{C1} in Figure~\ref{fig:C1}. On the left is the input C++ code, the middle of the figure depicts the extracted information as an in-memory hierarchical graph, and on the right is the generated TA model. In this example, Rex creates model nodes for the class \code{A}, function \code{updateX}, and variables \code{x}, \code{FA}, \code{FB}, and \code{GlobVar}\footnote{The names of the entities are simplified for this example to improve legibility. In practice, Rex creates long identifier names that capture the entity’s context (i.e.,
enclosing function, class, etc., up to and including filename).}.
Each \code{contain} edge corresponds to an entity declaration (e.g., class \code{A} contains the declaration of variable \code{x}). When one variable appears in an expression that is assigned to another variable (e.g., the use of  \code{GlobVar} in an assignment to variable \code{x} in \code{C1}), a \code{varWrite} edge is created from the used variable to the assigned variable (e.g., \code{varWrite GlobVar x}). The creation of the other edges follows the same pattern. Attributes of entities and relationships
are listed at the end of the TA model.
The attribute \code{PC} records presence conditions: any entity or relationship that is annotated with a PC attribute represents a fact that is conditionally present in the model, depending on the value of the PC's feature variables.
Thus, variability-aware Rex extracts a 150\% model representing facts from all products in the SPL, where conditional facts are annotated with their products' presence conditions.

In the last step of the extraction, the in-memory hierarchical graph is converted to a TA model, which is a textual representation of the graph. Because identifiers must be declared before being used in the model, the output lists: (\textit{i}) all graph nodes (representing entities), (\textit{ii}) all graph edges (representing relations), and finally (\textit{iii}) all attributes of nodes and edges. The tuple representations of nodes, edges, and attributes have the following structure: 

\smallskip

\code{\$INSTANCE $\langle$NODE\_ID$\rangle$ $\langle$NODE\_TYPE$\rangle$}

\code{$\langle$EDGE\_TYPE$\rangle$ $\langle$EDGE\_SOURCE$\rangle$ $\langle$EDGE\_TARGET$\rangle$}

\code{$\langle$ID$\rangle$ \{ $\langle$KEY$\rangle$ $=$ $\langle$VALUE$\rangle$ . . . \}}

\smallskip

Because of the nature of static analysis, the resulting model is an over-approximation of the program's actual set of facts: it may contain some facts that are infeasible (e.g., a function call in a conditional branch that never executes).

Facts are ported automatically to the TSV format using the \emph{ta2tsv} command-line tool that we wrote specifically for that purpose. In a TA model, presence conditions are not co-located with their associated facts, but rather are listed as attributes at the end of the file. 
Our \emph{ta2tsv} command-line tool associates the presence-condition attributes with their corresponding TSV records.

\subsection{Lifted Behaviour Alteration Analysis}
The original behaviour alteration analysis expects a model of a single product, not a product line.  A lifted behaviour alteration analysis operates on a 150\% factbase and is expected to compute the same set of the results that would be computed if the original analysis were applied to each product configuration; the lifted analysis is also expected to annotate each of the results with a presence condition indicating the set of products to which this particular result applies.

Instead of adapting the behaviour alteration analysis to become variability-aware, 
we decided to use the existing variability-aware Datalog engine \lsouffle~\cite{Shahin:2020a}. This way we were able to leverage all the optimizations in \lsouffle~to ensure the scalability of our solution to industrial-scale systems, and at the same time minimize the effort needed to build the components of the analysis pipeline.




\lsouffle~takes as input facts annotated with precedence conditions and infers additional facts based on a set of Datalog rules (the analysis logic in our case). Presence conditions of inferred facts are calculated as a part of the inference process. Those presence conditions are also checked for propositional satisfiability. Because an \term{unsatisfiable} PC indicates an empty set of products, inferred facts with unsatisfiable PCs are removed from the factbase.
\lsouffle~stores presence conditions as Binary Decision Diagrams (BDDs). This has the advantage of keeping a canonical representation of each presence condition, eliminating redundancies due to propositional logic identities (e.g., commutativity of conjunction and disjunction). CUDD~\cite{Somenzi:1998}, the BDD package used by \lsouffle, also caches BDDs, saving time on BDD construction. 

\begin{figure*}[t]
\begin{subfigure}[]{0.49\textwidth}
	\includegraphics[width=\textwidth]{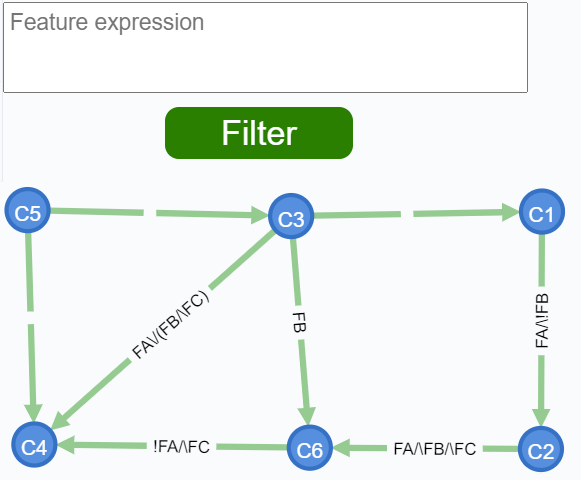}
	\caption{A visualization of component-interaction analysis results. A text box allows the user to filter the results by feature expression.}
	\label{fig:standardGraph}	
\end{subfigure}
\hfill
\begin{subfigure}[]{0.49\textwidth}
	\includegraphics[width=\textwidth]{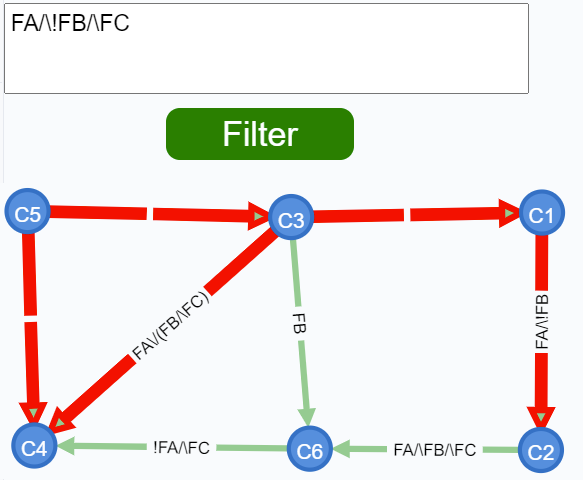}
	\caption{The user interface highlights (in red) the analysis results that are implied by the user-provided feature expression  $\FA~\land~!\FB~\land~\FC$.}
	\label{fig:highlightedGraph}	
\end{subfigure}
\caption{Visualization of variability-aware analysis results.}
\label{fig:graphViz}
\end{figure*}

\subsection{Interactive Visualization / Filtering}
As mentioned earlier, the results of a lifted behaviour alteration analysis are \textit{paths} in the factbase starting with a variable assignment, following data-flows from that variable to other variables whose values are affected by the first variable, and so on, ending with a function whose invocation may be influenced by one of the affected variables. 
If variable nodes and function nodes are abstracted to their component nodes, this same analysis result is represented as a path from the component containing the initial variable assignment to the component containing the influenced function call; the graph edges are labelled with the presence conditions of the products in which the behaviour alteration is possible.

This graph representation provides a useful overview of all possible interactions among components in the product line. However, the end user would benefit from 
the ability to explore and filter the results with respect to subsets of products of interest, and to compare how components interact in some products and not in others. Thus, we developed an interactive graph visualization interface that filters analysis results based on user input of a feature expression. 
Specifically, the user provides a feature expression representing a set of products of interest, and the visualizer highlights the graph edges whose presence conditions are implied by the provided feature expression. The visualization is implemented on top of the Neo4j Browser~\cite{neo4jBrowser}, which is the open-source user interface provided by the Neo4j graph database; it uses Logic Solver~\cite{logicSolver}, a boolean satisfiability solver, to reason whether feature expressions imply graph-edge presence conditions.



Figure~\ref{fig:standardGraph} shows the results of applying behaviour alteration analysis on an expanded version of the code presented in Figure~\ref{fig:splExample}; this example considers a system comprising 10 components that interact with one another. The visualization provides a textbox in the top left corner, in which the user can enter a feature expression, such as $\FA \land !\FB \land \FC$. The graph edges whose presence conditions are implied by the provided feature expression are highlighted in red (Figure~\ref{fig:highlightedGraph}). 

\section{Industrial SPL Examples}
\label{sec:caseStudy}

\newcommand{\featuresA}{$\sim$400}
\newcommand{\inputsA}{157303}
\newcommand{\inputPCsA}{698}
\newcommand{\inputPCPercA}{0.444}

\newcommand{\featuresB}{$\sim$500}
\newcommand{\inputsB}{225538}
\newcommand{\inputPCsB}{1070}
\newcommand{\inputPCPercB}{0.474}

\newcommand{\featuresC}{$\sim$900}
\newcommand{\inputsC}{215120}
\newcommand{\inputPCsC}{2125}
\newcommand{\inputPCPercC}{0.988}

\newcommand{\featuresD}{$\sim$600}
\newcommand{\inputsD}{228185}
\newcommand{\inputPCsD}{1148}
\newcommand{\inputPCPercD}{0.503}

\newcommand{\featuresE}{$\sim$600}
\newcommand{\inputsE}{227241}
\newcommand{\inputPCsE}{2078}
\newcommand{\inputPCPercE}{0.914}

\newcommand{\featuresF}{$\sim$500}
\newcommand{\inputsF}{226640}
\newcommand{\inputPCsF}{955}
\newcommand{\inputPCPercF}{0.421}

Our industrial study is performed on models extracted from six vehicle
controller product lines provided
by \GM, which are abstractly named  \EEightZ, \EEightE,..., \EZeroT to obfuscate sensitive industrial data.
Metrics on the sizes of all six product lines are shown in
Table~\ref{tbl:codeMetrics}. Some metrics (e.g., the number of features) are approximated to avoid revealing technical details.
For example, \EEightZ~has 5431 header (.h) files, with a total of 350,102 lines of code (LOC). It also has 5133 C language source files (.c), totalling 730,947 lines of code. Those C source files and header files implement approximately \featuresA~features.

\begin{table*}[t]
	\centering
	\caption{Size metrics for the six product lines analyzed. For each product line, we list the number of header files (.h) and C source files (.c), together with the total number of Lines of Code (LOC). The bottom half of the table lists the metrics of the extracted SPL models, 
	including the approximate number of features, the total number of extracted facts, the number of variational facts (i.e., those facts present only in a proper subset of an SPL's products), and the percentage of variational facts.}
	\begin{tabular}{l|cccccc} 
		\hline
		& \EEightZ & \EEightE & \ENine & \ENineN & \EZero & \EZeroT \\
		\hline
		(\texttt{.h}) Files & 5431 & 6277 & 4702 & 6292 & 5243 & 6115 \\
		(\texttt{.h}) LOC & 350,102 & 570,174 & 285,132 & 586,985 & 337,946 & 572,851 \\
		(\texttt{.c}) Files & 5133 & 6826 & 4300 & 6943 & 4981 & 6464 \\
		(\texttt{.c}) LOC & 730,947 & 1,016,063& 750,000 & 979,466 & 752,669 & 1,088,811 \\
	    \hline
	    Features & \featuresA & \featuresB & \featuresC & \featuresD & \featuresE & \featuresF \\
		Input facts & \inputsA & \inputsB & \inputsC & \inputsD & \inputsE & \inputsF \\
		Input facts with explicit PCs & \inputPCsA & \inputPCsB & \inputPCsC & \inputPCsD & \inputPCsE & \inputPCsF \\
		Input facts with explicit PCs (\%) & \inputPCPercA\% & \inputPCPercB\% & \inputPCPercC\% & \inputPCPercD\% & \inputPCPercE\% & \inputPCPercF\%  \\
		\hline
	\end{tabular}
	\label{tbl:codeMetrics}
\end{table*}

\GM's controller code encodes inclusion or exclusion of features using \term{calibration parameters}.  
Truth values of calibration parameters encode variability
in a way that allows for easy variation as to which features are enabled in any particular 
vehicle.
The calibration parameters in \GM's code are encoded
as global constants of enumerated types (\texttt{enum}) or boolean type (\texttt{bool}).
As these constants represent calibration parameters, their values are defined at deployment time during vehicle manufacturing~\cite{Young:2017}. 
Such an encoding of variability means 
that the source code includes all of the code relevant to all features. Thus, each controller code-base is a 150\% representation of the controller's SPL, and an individual controller product is configured by setting the values of these calibration parameters. The 
SPL in Figure~\ref{fig:splExample} is a demonstration of this kind of 
variability representation, where the \texttt{bool} variables \texttt{FA} and \texttt{FB} are examples of calibration parameters. 

The software of a vehicle has many variation points and thus configuration
involves many calibration parameters~\cite{Young:2017}. 
In our SPL examples, the code has several hundred calibration parameters (features in Table~\ref{tbl:codeMetrics}) in each of the controllers.
Because the number of possible products is exponential in the number of calibration parameters, the
large number of calibration parameters makes analyzing individual products infeasible.

As the behaviour alteration analysis targets interactions between functions in distinct components,
\GM{} has shared with us a high-level decomposition of their code into components.  These components
comprise several source-code files with shared data.
Some variables or constants are deemed internal to a component, and 
read or write access to that 
data from another component indicates a potential error.
Facts in the factbase are annotated with their component name, which is
used in the behaviour alteration analysis (see Figure~\ref{fig:HS5}) to eliminate
intra-component results. 


\section{Applying Analysis to the Industrial Examples}
\label{sec:evaluation}

\newcommand{\timeBaseA}{4.093}
\newcommand{\outputsBaseA}{125837}
\newcommand{\uniquePCsA}{167}
\newcommand{\timeA}{6.149}
\newcommand{\outputsA}{125819}
\newcommand{\outputPCsA}{444}
\newcommand{\outputPCPercA}{0.353}
\newcommand{\overheadA}{50.232}
\newcommand{\unsatsA}{18}
\newcommand{\unsatPercA}{0.014}

\newcommand{\timeBaseB}{5.296}
\newcommand{\outputsBaseB}{173687}
\newcommand{\uniquePCsB}{242}
\newcommand{\timeB}{6.737}
\newcommand{\outputsB}{173685}
\newcommand{\outputPCsB}{624}
\newcommand{\outputPCPercB}{0.359}
\newcommand{\overheadB}{27.209}
\newcommand{\unsatsB}{2}
\newcommand{\unsatPercB}{0.001}

\newcommand{\timeBaseC}{18.222}
\newcommand{\outputsBaseC}{394770}
\newcommand{\uniquePCsC}{680}
\newcommand{\timeC}{19.569}
\newcommand{\outputsC}{394712}
\newcommand{\outputPCsC}{3269}
\newcommand{\outputPCPercC}{0.828}
\newcommand{\overheadC}{7.392}
\newcommand{\unsatsC}{58}
\newcommand{\unsatPercC}{0.015}

\newcommand{\timeBaseD}{4.934}
\newcommand{\outputsBaseD}{162898}
\newcommand{\uniquePCsD}{259}
\newcommand{\timeD}{7.857}
\newcommand{\outputsD}{162897}
\newcommand{\outputPCsD}{702}
\newcommand{\outputPCPercD}{0.431}
\newcommand{\overheadD}{59.245}
\newcommand{\unsatsD}{1}
\newcommand{\unsatPercD}{0.001}

\newcommand{\timeBaseE}{18.295}
\newcommand{\outputsBaseE}{429125}
\newcommand{\uniquePCsE}{517}
\newcommand{\timeE}{19.274}
\newcommand{\outputsE}{427569}
\newcommand{\outputPCsE}{18662}
\newcommand{\outputPCPercE}{4.365}
\newcommand{\overheadE}{5.351}
\newcommand{\unsatsE}{1556}
\newcommand{\unsatPercE}{0.363}

\newcommand{\timeBaseF}{4.992}
\newcommand{\outputsBaseF}{173784}
\newcommand{\uniquePCsF}{233}
\newcommand{\timeF}{6.720}
\newcommand{\outputsF}{173782}
\newcommand{\outputPCsF}{880}
\newcommand{\outputPCPercF}{0.506}
\newcommand{\overheadF}{34.613}
\newcommand{\unsatsF}{2}
\newcommand{\unsatPercF}{0.001}

One of the primary goals of this project was to validate that the variability-aware Datalog analysis approach~\cite{Shahin:2019} is scalable to real-life industrial SPLs. We informally define scalability as having a marginal performance overhead compared to analyzing the \onefifty~of the SPL, which implicitly means having an exponential speedup compared to product-based analysis of each single product individually.

The bottom half of Table~\ref{tbl:codeMetrics} lists size metrics for the extracted  models of the industrial example SPLs. In terms of the number of extracted facts, the sizes of the models range from \inputsA~(\EEightZ) to \inputsD~(\ENineN). The numbers of SPL features in the subject SPLs (approximate numbers, as requested by \GM) range from
\featuresA~(\EEightZ) to \featuresC~(\ENine). These feature counts are an order of magnitude higher than the feature counts used in the evaluation of variability-aware Datalog conducted in~\cite{Shahin:2019}. 
The last two rows in the table list, for each product line, the number of variational facts (i.e., those facts with explicit presence conditions) and their percentage with respect to the total number of facts extracted from that product line. 
For example, \inputsA~facts were extracted from \EEightZ, of which \inputPCsA~are variational (i.e., they are present in a proper subset of products); the percentage of \EEightZ~facts that are variational is \inputPCPercA\%. 

\begin{table*}[tbp]
	\centering
	\caption{Results of analyzing the models of the six product lines.}
	\begin{tabular}{l|cccccc} 
		\hline
		& \EEightZ & \EEightE & \ENine & \ENineN & \EZero & \EZeroT \\
		\hline
		\onefifty~time & \timeBaseA~sec. & \timeBaseB~sec. & \timeBaseC~sec. & \timeBaseD~sec. & \timeBaseE~sec. & \timeBaseF~sec.\\
		\onefifty~output facts & \outputsBaseA & \outputsBaseB & \outputsBaseC & \outputsBaseD & \outputsBaseE & \outputsBaseF \\ 
		\hline
		Unique PCs & \uniquePCsA & \uniquePCsB & \uniquePCsC & \uniquePCsD & \uniquePCsE & \uniquePCsF \\
		Lifted analysis time & \timeA~sec. & \timeB~sec. & \timeC~sec. & \timeD~sec. & \timeE~sec. & \timeF~sec. \\
		Output facts & \outputsA & \outputsB & \outputsC & \outputsD & \outputsE & \outputsF \\
		Output facts with PCs & \outputPCsA & \outputPCsB & \outputPCsC & \outputPCsD & \outputPCsE & \outputPCsF \\
		Output facts with PCs (\%) & \outputPCPercA\% & \outputPCPercB\% & \outputPCPercC\% & \outputPCPercD\% & \outputPCPercE\% & \outputPCPercF\% \\
		\hline
		Time overhead & \overheadA\% & \overheadB\% & \overheadC\% & \overheadD\% & \overheadE\% & \overheadF\% \\
		Unsat output facts & \unsatsA & \unsatsB & \unsatsC & \unsatsD & \unsatsE & \unsatsF \\
		Unsat output facts (\%) & \unsatPercA\% & \unsatPercB\% & \unsatPercC\% & \unsatPercD\% & \unsatPercE\% & \unsatPercF\% \\
		\hline
	\end{tabular}
	\label{tbl:results}
\end{table*}

\forreview{ 
For each controller SPL, we used variability-aware Rex to extract automatically a \textit{\onefifty~}(i.e., a model representing a single product with all features present) and an SPL model, with feature variability represented as presence-condition annotations on facts.}
We translated the extracted facts into Datalog facts. Within \lsouffle, we applied the original behaviour-alteration analysis (expressed as Datalog rules) to each subject's \onefifty~and applied the variability-aware behaviour-alteration analysis (also expressed in Datalog) to each subject's SPL model, repeating the analyses on each model five times and reporting the average execution time after excluding the minimum and maximum times.  

Table~\ref{tbl:results} summarizes the results of the experiments. For example, column \EEightZ~reports on the results for controller \EEightZ): analysis of the \onefifty~takes \timeBaseA~seconds, whereas variability-aware analysis of the SPL model takes \timeA~seconds, with a time overhead of \overheadA\%. 
Analysis of the \onefifty~produces \outputsBaseA~output facts, whereas variability-aware analysis of the SPL model produces \unsatsA~fewer output facts (i.e., \outputsA~facts). Those \unsatsA~facts (representing \unsatPercA\% of \EEightZ's output facts) belong to none of the product variants of \EEightZ; they have unsatisfiable presence conditions and thus are eliminated as part of the variability-aware analysis that reasons about presence conditions.
Analysis of the SPL model produces \outputPCsA~output facts that have explicit presence conditions (i.e., these facts are present only in a proper subset of \EEightZ's product variants), which is \outputPCPercA\% of the output facts; only \uniquePCsA~unique presence conditions are used in these annotations.

The execution times of the analyses on \onefifty~models range from \timeBaseA~seconds (\EEightZ) to \timeBaseE~seconds (\EZero); whereas the execution times of the analyses on SPL models range from \timeA~seconds (\EEightZ) to \timeC~seconds (\ENine), with overheads that range from \overheadE\% (\EZero) to \overheadD\% (\ENineN). Recall that the cost of product-based analysis (where each product of an SPL is analyzed separately) grows exponentially with the number of features~\cite{Liebig:2013}. Thus, it is noteworthy that the execution-time overhead of our variability-aware behaviour-alteration analysis does not seem to correlate with the number of SPL features. The marginal overheads incurred can be considered \emph{very acceptable}, at least in cases like our industry examples, where a system has hundreds of features but sparse variability in terms of the percentage of facts annotated with presence conditions.

In addition to execution time, we also measured the number of facts generated by the analyses, including all intermediate facts generated during inference. As mentioned above, the reason there is a difference in the number of facts inferred by the two analyses is that the variability-aware analysis excludes facts that have unsatisfiable presence conditions, in order to improve the accuracy of the analysis results; whereas in the analysis of a \onefifty, all inferred facts are deemed to be feasible.

With the exception of the \EZero~controller, the variability-aware analysis on each controller's SPL model inferred less than 0.1\% fewer facts than the analysis on that controller's \onefifty.
Even for the \EZero~controller, the reduction in the number of facts inferred by the variability-aware analysis is only about 0.36\%.

\forreview{
}

We measured also the total number of unique presence conditions computed during the inference process\footnote{This measurement was aided by the fact that the presence conditions have canonical BDD representations.}.

To our surprise, the number of unique presence conditions were smaller than the number of features for each of the six SPLs -- which is far fewer than the number of possible combinations of features. Taking a further look at the presence conditions, we found out that many features always appear together in a presence condition. This kind of feature correlation is not uncommon in SPLs~\cite{Apel:2011}.

In summary, with a performance overhead of only 5-59\% compared to the analysis of the single product with all features present (the \onefifty), our evaluation shows that variability-aware analysis scales to large-scale industrial software product lines with hundreds of features.

A secondary product of our work is the use of graph visualization and interactive techniques to enable the engineer to explore and filter the 
analysis results by feature expression. We hypothesize that colouring the subset of edges associated with a user-provided feature expression increases the readability of the analysis results by highlighting the interactions that match the user's filter while still offering a view of the facts' context.
This hypothesis must still be tested 
by conducting user studies that evaluate the filtering and highlighting of analysis results in different scenarios.  

Lastly, although we have not had an opportunity to seek detailed feedback from GM engineers (e.g., via a qualitative study), our presentations to GM engineers have elicited expressions of overall interest in all aspects of the tool chain. In particular, they considered the facts that are collected and their variabilities to be extensive, and they see the use of a query language for posing specific and ad-hoc queries to the fact base to be potentially quite powerful. They also felt that, with more specific facts about the software and the product-line, more accurate facts could be derived from the tool chain.


\section{Lessons Learned}
\label{sec:lessons}
In this section, we reflect on some of the lessons learned by conducting this project.

\subsection{Scalability of Lifted Analysis}
In theory, the complexity of software product line analysis is expected to grow with respect to the number of product line features~\cite{Liebig:2013}. Product variants compose features together, thus the number of product variants typically grows exponentially with the number of features. The idea behind lifting analyses to product lines is to leverage the commonality among different product variants as much as possible to keep the cost of product line analysis reasonable, as opposed to enumerating and analyzing each product variant by itself, which is intractable in most practical cases.

The product lines we analyzed in this study have hundreds of features each, which means that enumerating each product is not an option. The variability-aware overhead reported for \lsouffle~in earlier work~\cite{Shahin:2019} is marginal, but that was reported for relatively small benchmarks (none of them of industrial scale) of only tens of features each. Results presented in Section~\ref{sec:evaluation} show that the performance overhead of full product line analysis using \lsouffle~is still marginal for industrial product lines, with hundreds of features.

Looking further into the results, the performance overhead does not seem to correlate with the size of the code-base, the size of the extracted model (number of facts), or the number of features of the SPL. This can be explained by differences between the subject SPLs with respect to the code patterns directly relevant to the particular analysis applied. Also measuring the unique number of presence conditions generated throughout the analysis sheds some light on how some features are tightly coupled in industrial product lines, causing the effective complexity of the analysis to be lower than what might be perceived given the number of features.

\subsection{Variability Encoding}

\lsouffle~can only handle binary features; that is, a feature can be either present or absent. However, the SPLs we analyzed in this project also encode sets of mutually exclusive features using C-language \code{enum} data types. For example, if \code{Feat0}, \code{Feat1}, \code{Feat2}, and \code{Feat3} is a set of four mutually exclusive features, it is a common C-language idiom to encapsulate them in an enumerated data type:
\lstset{style=GrokStyle}
\begin{lstlisting}[language=c]
enum FeatSet {
    Feat0,
    Feat1,
    Feat2,
    Feat3
};
\end{lstlisting}

\noindent Enumerated data types in C are integral types, allowing the use of mathematical integer operators (e.g., addition, conjunction, bit-wise disjunction) and comparison operators on their values. We came across cases where presence conditions included comparison operators on values of enumerated data types, and we had to abstract those predicates into propositional symbols.
For example, if \code{x} is a constant of type \code{FeatSet}, then the expression \code{x < Feat2} is a logically valid presence condition, but is not acceptable in \lsouffle. We apply a syntactic transformation for these kinds of expressions, turning the above expression into a boolean \code{x\_LT\_Feat2}, where the \code{\_LT\_} sub-string stands for less-than. We use similar substitutions for other comparison operators.
 
The fact that the four features belonging to the \code{FeatSet} are mutually exclusive can be then added to the feature model of the product line. The fragment of the feature model representing this property for \code{FeatSet} is:
\[\vspace{-0.1in} \lnot~(\code{Feat0}~\land~\code{Feat1})~\land~\lnot~(\code{Feat0}~\land~\code{Feat2})~\land \]
\[\vspace{-0.1in} \lnot~(\code{Feat0}~\land~\code{Feat3})~\land~\lnot(\code{Feat1}~\land~\code{Feat2})~\land\]
\[\lnot(\code{Feat1}~\land~\code{Feat3})~\land~\lnot(\code{Feat2}~\land~\code{Feat3}) \]

If a feature from \code{FeatSet} is mandatory, we also need to add the disjunction of all four features to the feature model:
\[(\code{Feat0}~\lor~\code{Feat1}~\lor~\code{Feat2}~\lor~\code{Feat3})\]

\subsection{Variability Annotation}

Different techniques have been used to annotate segments of source-code with feature expressions, effectively deciding which pieces of code belong to which features. For example, CIDE~\cite{Kastner:2009} is a colour-based tool that highlights segments of code with different colours, each of which represents a feature. The most commonly used annotation mechanism in industrial product lines is the C Pre-Processor (CPP)~\cite{Ernst:2002, Liebig:2010}. The CPP provides a high degree of flexibility when annotating source code, allowing for lexical rather than syntactic annotation. This means that any sequence of lexemes (tokens), even if the sequence by itself is not syntactically valid, can be assigned a presence condition. As a result, the \onefifty~of an SPL annotated with CPP directives is typically not syntactically well-formed, requiring variability-aware parsing~\cite{Gazzillo:2012,Kastner:2011}.

The product lines from \GM, however, use a different annotation mechanism. C-language constants (following a naming convention) are used within the source code to indicate features. Those constants are assigned values as a part of the product configuration process. Feature-specific code is thus enclosed within C-language conditional statements, relying on the compiler to evaluate the compile-time constants at compile time and to eliminate dead-code corresponding to features not included in the product being built.

This annotation technique has two direct consequences. First, while it is less flexible than CPP directives, it does not require
variability-aware parsing because the entire product line is a syntactically well-formed C-program. Secondly, existing analysis tools can be applied to the entire product line, in the same way as regular parsers can be applied to it. The downside is that each result of a given analysis is not labeled with the distinct set of products to which it applies. This draws a clear distinction between analyzing the \onefifty~of a product line, in the case where it is well-formed and readable by an analysis tool, and variability-aware analysis, where both inputs and outputs of the analysis need to be appropriately annotated.

An indirect consequence of the annotation technique used by \GM~is the possibility of filtering analysis results through 
user-provided feature expression of interest and presenting only facts with satisfying presence conditions. This capability has the potential to improve the readability of the data and support the experience of the user visually inspecting the analysis results.

\section{Related Work}
\label{sec:related}

\textit{\textbf{Variability-Aware Analysis}.}
Different kinds of source-code analyses have been re-implemented to be variability aware~\cite{Thum:2014}. For example, the TypeChef project~\cite{Kastner:2011,Kastner:2012} implements variability-aware parsing~\cite{Kastner:2011} and type checkers~\cite{Kastner:2012} for Java and C. The SuperC project~\cite{Gazzillo:2012} is another C language variability-aware parser. With respect to model-based analyses, the Henshin~\cite{Arendt:2010} graph-transformation engine was lifted to support product lines of graphs~\cite{Salay:2014}. These lifted analyses were written from scratch, without reusing any components from their respective product-based analyses. Our approach, on the other hand, lifts an entire class of product-based analyses written as Datalog rules, by lifting their inference engine (and extracting presence conditions together with facts).

SPL\textsuperscript{Lift}~\cite{Bodden:2013} extends IFDS~\cite{Reps:1995} data-flow analyses to product lines. Model checkers based on Featured Transition Systems~\cite{Classen:2013} check temporal properties of transition-system models where transitions can be labeled by presence conditions. Both of these SPL analyses use almost the same single-product analyses on a lifted data representation. At a high level, our approach is similar in the sense that the logic of the original analysis is preserved, and only data is augmented with presence conditions. Still, our approach is unique because we do not touch any of the Datalog rules comprising the
analysis logic itself.

Lifting query languages used to implement analyses instead of lifting a single analysis is the approach we are using in this paper. Particularly, we use a variability-aware Datalog engine~\cite{Shahin:2020a} that implicitly lifts analyses written in Datalog~\cite{Shahin:2019}. This approach has also been recently extended to lift analyses written in more expressive, Turing-complete languages~\cite{Shahin:2020b}.

\textit{\textbf{Variability-aware Visualization.}} Colour is used effectively in SPL visualization to represent traceability links between source code and feature models. Tools like CIDE~\cite{kastner2009guaranteeing}, FeatureMapper~\cite{Heidenreich-VISPLE08}, fmp2rsm~\cite{czarnecki2006verifying}, and FeatureVISU~\cite{Apel:2011}
enable colouring of model entities or source-code fragments according to their association with a set of features that the user selects. Our visualization can similarly aid a user in visualizing the results of a variability-aware analysis that apply to a subset of the SPL's products.
Visualization tools that employ interactive techniques, such as detail-on-demand and highlighting, are proven to contribute to the engineer's comprehension of a product line and to their productivity in modifying the feature configurations~\cite{asadi2016effects}. 

The visualization presented in~\cite{loesch2007optimization} similarly supports an analysis of the feature configuration of a software product line.  The authors use Formal Concept Analysis to identify obsolete variable features to optimize the configuration of the product line. Their graph visualization explores the spatial distribution of the nodes and their size to respectively encode the difference between objects and attributes, and the number of feature variables associated with each node. Our work differs in terms of the goal of the analysis and data encoding provided by the visualization. 

Work that is closer to ours include the visualizations provided by VISIT-FC~\cite{botterweck2008visual}, which support the understanding of possible consequences of the engineer's decision. The tool provides an interactive view connecting three models (decision, feature, and component models) where the user can select features and visualize the decisions and components related to their selection and the relations between them. The traceability is visualized by explicit links connecting the models' components. The highlighting of those links is performed by colouring all non-relevant entities in gray, colouring only the  information relevant to the engineer. 
Our visualization focuses on filtering and highlighting of variability-aware analysis results with respect to how the results apply to particular product sets, rather than filtering and highlighting subsets of a variable system. We recognize that both types of visualization focus on feature- and product-specific aspects of some base representation that exhibits variability, and techniques developed to visualize slices of systems-with-variability could be used to visualize and highlight slices of analysis results-with-variability.
\label{sec:conclusion}
\section{Conclusion and Future Work}

In this paper, we presented an industrial study of applying a declarative source-code analysis to relational models of annotative Software Product Lines (SPLs). We integrated source-code fact extraction and a variability-aware Datalog engine from two prior projects~\cite{Shahin:2019,Muscedere:2019}, implementing an analysis pipeline. In addition to adapter components between pieces coming from different projects, we enhanced the fact extraction to be variability-aware and added a result-filtering and visualization module for the interactive inspection of results.

We applied the pipeline to a component-interaction behaviour-alteration analysis of models of six automotive controller SPLs from \GM, each with hundreds of product line features. Our results demonstrate the scalability of our variability-aware analysis approach to real-life industrial SPLs. Analyzing the whole SPL, with presence conditions that relate source-code facts to their associated products, was only 5-59\% slower than analyzing the entire factbase including all features with no presence conditions. Our interactive visualization module allows users to filter the analysis results for a subset of products, allowing for a finer-grained, product-level inspection of results when needed.

For future work, we plan to integrate our analysis pipeline more tightly to produce a single tool that takes SPL code as input and provides an interactive user interface for inspecting results. We are also in discussions with \GM~to apply the pipeline to other analyses and to more SPLs. In addition, since our pipeline is analysis-agnostic, we are also in the process of identifying other analyses that might be of value to \GM~and whether they can be implemented in Datalog.
We also aim to validate the usability of our visualization module for end users.  
\bibliographystyle{IEEEtran}
\bibliography{spl,datalog,viz,factExtract}
\end{document}